
\input harvmac.tex
\Title{CTP/TAMU-18/93}{A Comment on String Solitons\footnote{$^\dagger$}
{Work supported in part by NSF grant PHY-9106593.}}

\centerline{
Ramzi R.~ Khuri\footnote{$^*$}{Supported by a World Laboratory Fellowship.}}
\bigskip\centerline{Center for Theoretical Physics}
\centerline{Texas A\&M University}\centerline{College Station, TX 77843}

\vskip .3in
We derive an exact string-like soliton solution of $D=10$ heterotic string
theory. The solution possesses $SU(2)\times SU(2)$ instanton structure in the
eight-dimensional space transverse to the worldsheet of the soliton.

\Date{4/93}


\def\sqr#1#2{{\vbox{\hrule height.#2pt\hbox{\vrule width
.#2pt height#1pt \kern#1pt\vrule width.#2pt}\hrule height.#2pt}}}
\def\Box{\mathchoice\sqr64\sqr64\sqr{4.2}3\sqr33}

\def\gctwo{\hat R^i{}_{jkl}}
\def\gcone{\hat R^\alpha{}_{\beta\gamma\lambda}}

\def\met {g_{\mu\nu}}

\lref\inst{R.~R.~Khuri, Phys. Lett. {\bf B259} (1991) 261.}

\lref\monin{R.~R.~Khuri, Phys. Rev. {\bf D46} (1992) 4526.}

\lref\chsone{C.~G.~Callan, J.~A.~Harvey and A.~Strominger, Nucl. Phys.
{\bf B359} (1991) 611.}

\lref\strom{A.~Strominger, Nucl. Phys. {\bf B343} (1990) 167.}

\lref\fbrane{M.~J.~Duff and J.~X.~Lu, Nucl. Phys. {\bf B354} (1991) 141.}

\lref\duality{M.~J.~Duff and J.~X.~Lu, Nucl. Phys. {\bf B354} (1991) 129.}

\lref\rey{S.~J.~Rey, Phys. Rev. {\bf D43} (1991) 526.}

\lref\chad{J.~M.~Charap and M.~J.~Duff, Phys. Lett. {\bf B69} (1977) 445.}

\lref\dfluhs{M.~J.~Duff and J.~X.~Lu, Phys. Rev. Lett. {\bf 66} (1991) 1402.}

\lref\koun{C.~Kounnas, to appear in {\it Proceedings of INFN Eloisatron
Project, 26th Workshop: ``From Superstrings to Supergravity", Erice, Italy,
Dec. 5-12, 1992, World Scientific, ed. M.~Duff, S.~Ferrara and R.~Khuri.}}

In \chsone, an exact multi-fivebrane soliton solution of heterotic string
theory was presented. This solution represented an exact extension of the
tree-level supersymmetric multi-fivebrane solutions of \refs{\strom,\fbrane}.
For this class of fivebrane solutions, the generalized curvature incorporating
the axionic field strength possesses a (anti) self-dual structure
\refs{\inst,\monin}\
and is referred to as an ``axionic instanton" (see \rey\ and references
therein). Exactness is shown for the heterotic solution based on algebraic
effective action arguments and $(4,4)$ worldsheet supersymmetry \chsone.
The gauge sector of the heterotic solution possesses $SU(2)$ instanton
structure in the four-dimensional space transverse to the fivebrane.
In more recent work, Kounnas \koun\ described a method of obtaining
string solutions with non-trivial backgrounds by using $N=4$ superconformal
building blocks with $\hat c=4$. In particular, he proposed the existence
of an exact solution with $SU(2)\times SU(2)$ instanton structure.

In this paper we obtain an explicit space-time background corresponding
to Kounnas' conformal field theory by constructing an exact string-like
solution of $D=10$ heterotic string theory from a modification of the fivebrane
ansatz. In the eight-dimensional space transverse to the string, the solution
contains two independent $SU(2)$ instantons each embedded in a separate $SO(4)$
subgroup of the gauge group. The arguments demonstrating exactness of this
solution follow those of \chsone.

The tree-level supersymmetric vacuum equations for the heterotic string are
given by
\eqn\sseq{\eqalign{\delta\psi_M&=\left(\nabla_M-{\textstyle {1\over 4}}H_{MAB}
\Gamma^{AB}\right)\epsilon=0,\cr
\delta\lambda&=\left(\Gamma^A\partial_A\phi-{\textstyle{1\over 6}}
H_{ABC}\Gamma^{ABC}\right)\epsilon=0,\cr
\delta\chi&=F_{AB}\Gamma^{AB}\epsilon=0, \cr}}
where $\psi_M,\ \lambda$ and $\chi$ are the gravitino, dilatino and gaugino
fields. The Bianchi identity is given by
\eqn\bianchi{dH={\alpha'\over 4} \left({\rm tr} R\wedge R-{\rm tr}
F\wedge F\right).}
The $(9+1)$-dimensional Majorana-Weyl fermions decompose down to
chiral spinors according to
$SO(9,1)\supset SO(1,1) \otimes SO(4) \otimes SO(4)$ for
the $M^{9,1}\to M^{1,1}\times M^4\times M^4$ decomposition. The ansatz
\eqn\bifbsol{\eqalign{\phi&=\phi_1 + \phi_2,\cr
\met&=e^{2\phi_1}\delta_{\mu\nu}\qquad \mu,\nu=2,3,4,5,\cr
g_{mn}&=e^{2\phi_2}\delta_{mn}\qquad m,n=6,7,8,9,\cr
g_{ab}&=\eta_{ab}\qquad\quad   a,b=0,1,\cr
H_{\mu\nu\lambda}&=\pm 2\epsilon_{\mu\nu\lambda\sigma}\partial^\sigma\phi
\qquad \mu,\nu,\lambda,\sigma=2,3,4,5,\cr
H_{mnp}&=\pm 2\epsilon_{mnpk}\partial^k\phi
\qquad m,n,p,k=6,7,8,9\cr}}
with constant chiral spinors
$\epsilon_\pm=\epsilon_2 \otimes \eta_4 \otimes \eta_4'$ solves the
supersymmetry equations with zero background fermi fields provided the YM gauge
field satisfies the instanton (anti) self-duality condition
\eqn\ymin{\eqalign{F_{\mu\nu}&=\pm {1\over
2}\epsilon_{\mu\nu}{}^{\lambda\sigma}
F_{\lambda\sigma},\qquad \mu,\nu,\lambda,\sigma=2,3,4,5\cr
F_{mn}&=\pm {1\over 2}\epsilon_{mn}{}^{pk} F_{pk},\qquad m,n,p,k=6,7,8,9.\cr}}
The chiralities of the spinors $\epsilon_2$, $\eta_4$ and $\eta_4'$
are correlated by
\eqn\chicor{(1 \mp \gamma_3)\epsilon_2 = (1 \mp \gamma_5)\eta_4
= (1 \mp \gamma_5)\eta_4'=0,}
so that three-quarters of the spacetime supersymmetries are broken.
An exact solution is obtained as follows. Define a generalized connection by
\eqn\gcon{\Omega^{AB}_{\pm M}=\omega^{AB}_M\pm H^{AB}_M }
in an $SU(2) \times SU(2)$ subgroup of the gauge group, and equate it
to the gauge connection $A_M$ \chad\ for $M=2,3,4,5,6,7,8,9$ so that $dH=0$ and
the corresponding curvature $R(\Omega_{\pm})$ cancels against the Yang-Mills
field strength $F$ in both subspaces $(2345)$ and $(6789)$.
For $e^{-2\phi_1}\Box\ e^{2\phi_1}=e^{-2\phi_2}\Box\ e^{2\phi_2}=0$, the
curvature of the generalized connection can be written in covariant form
\refs{\inst,\monin}
\eqn\gcurvone{\eqalign{\gcone&=
\delta_{\alpha\lambda}\nabla_\gamma\nabla_\beta\phi_1
-\delta_{\alpha\gamma}\nabla_\lambda\nabla_\beta\phi_1
+\delta_{\beta\gamma}\nabla_\lambda\nabla_\alpha\phi_1
-\delta_{\beta\lambda}\nabla_\gamma\nabla_\alpha\phi_1 \cr
&\pm\epsilon_{\alpha\beta\gamma\mu}\nabla_\lambda\nabla_\mu\phi_1
\mp\epsilon_{\alpha\beta\lambda\mu}\nabla_\gamma\nabla_\mu\phi_1,\cr}}
where $\alpha,\beta,\gamma,\lambda,\mu=2,3,4,5$ and
\eqn\gcurvtwo{\eqalign{\gctwo&=\delta_{il}\nabla_k\nabla_j\phi_2
-\delta_{ik}\nabla_l\nabla_j\phi_2+\delta_{jk}\nabla_l\nabla_i\phi_2
-\delta_{jl}\nabla_k\nabla_i\phi_2 \cr
&\pm\epsilon_{ijkm}\nabla_l\nabla_m\phi_2
\mp\epsilon_{ijlm}\nabla_k\nabla_m\phi_2,\cr}}
where $i,j,k,l,m=6,7,8,9$. It easily follows that
\eqn\axinone{\gcone=\mp {1\over 2} \epsilon_{\gamma\lambda}{}^{\mu\nu}
\hat R^\alpha_{\beta\mu\nu},}
and
\eqn\axintwo{\gctwo=\mp {1\over 2} \epsilon_{kl}{}^{mn}\hat R^i_{jmn},}
from which it follows that both $F$ and $R$ are (anti) self-dual in both
four-dimensional subspaces. This solution becomes exact since
$A_M=\Omega_{\pm M}$ implies that all the higher order corrections vanish.
Both the algebraic effective action arguments and the $(4,4)$ worldsheet
supersymmetry arguments of \chsone\ can be used in essentially
the same manner to demonstrate exactness of the string solution. The explicit
solution for $\phi_1$ and $\phi_2$ in \bifbsol\ is given by
\eqn\multin{\eqalign{e^{2\phi_1}&=e^{2\phi_{1_0}}\left(1+
\sum_{i=1}^N{\rho_i^2\over |\vec x - \vec a_i|^2}\right),\cr
e^{2\phi_2}&=e^{2\phi_{2_0}}\left(1+
\sum_{j=1}^{M}{\lambda_j^2\over |\vec y - \vec b_j|^2}\right),\cr}}
where $\vec x$ and $\vec a_i$ are four-vectors and $\rho_i$ instanton scale
sizes in the space $(2345)$, and $\vec y$ and $\vec b_j$ are four-vectors
and $\lambda_j$ instanton scale sizes in the space $(6789)$. Axion charge
quantization then requires that $\rho_i^2=e^{-2\phi_{1_0}}n_i\alpha'$ and
$\lambda_j^2=e^{-2\phi_{2_0}}m_j\alpha'$, where $n_i$ and $m_j$ are integers.
Note that for $N=0$ or $M=0$ we recover the solution of \chsone.
It is interesting to note that both the charge $Q_2=-1/2 \int_{S^7}{}^\ast H$
and the mass per unit length ${\cal M}_2$ of the infinite string diverge.
By contrast, all classes of fivebrane solutions have finite charge and mass
per unit length as a result of the preservation of half the spacetime
supersymmetries and the saturation of a Bogmol'nyi bound. The fact that
three-quarters of the spacetime supersymmetries are broken for this solution
means that the saturation of the Bogomol'nyi bound is no longer guaranteed,
but it is unclear as to whether this would necessarily imply infinite mass
per unit length for the string. It would be interesting to see whether any
finite mass per unit length analogs of this solution exist, especially in the
context of the conjectured dual theory of fundamental fivebranes \duality.
Another interesting point is that the $D=8$ instanton number $N_8$ for this
string solution is in general nonzero for gauge group $E_8\times E_8$
($N_8=NM$, where $N$
and $M$ are the $D=4$ instanton numbers in the $(2345)$ and $(6789)$ spaces
respectively), since in this case $({\rm Tr F}^2)^2$ is nonvanishing. This is
to be contrasted with the zero $D=8$ instanton number found for the string
soliton solution of Duff and Lu \dfluhs. \footnote{$^\dagger$}{This was pointed
out to me by Mike Duff.}

\bigbreak\bigskip\bigskip\centerline{{\bf Acknowledgements}}\nobreak
I would like to thank Mike Duff for suggesting this problem and for
helpful discussions.

\vfil\eject
\listrefs
\bye